\documentclass[fleqn,10pt]{wlscirep}
\title{Explicit size distributions of failure cascades redefine systemic risk on finite networks}

\author[1,*]{Rebekka Burkholz}
\author[2,3]{Hans J. Herrmann}
\author[4]{Frank Schweitzer}
\affil[1]{ETH Zurich, Computer Science, Zurich, 8092, Switzerland}
\affil[2]{ETH Zurich, Computational  Physics  for  Engineering  Materials,  IfB, Zurich, 8093, Switzerland}
\affil[3]{Universidade  Federal  do  Cear\'{a}, F\'{i}sica, Fortaleza Cear\'{a},  60451-970, Brazil}
\affil[4]{ETH Zurich, Management, Technology and Economics, Zurich, 8092, Switzerland}

\affil[*]{rburkholz@ethz.ch}


\begin{abstract}
How big is the risk that a few initial failures of nodes in a network amplify to large cascades that span a substantial share of all nodes?  
Predicting the final cascade size is critical to ensure the functioning of a system as a whole. 
Yet, this task is hampered by uncertain or changing parameters and missing information. 
In infinitely large networks, the average cascade size can often be well estimated by established approaches building on local tree approximations and mean field approximations. 
Yet, as we demonstrate, in finite networks, this average does not even need to be a likely outcome.
Instead, we find broad and even bimodal cascade size distributions. 
This phenomenon persists for system sizes up to $10^{7}$ and different cascade models, i.e. it is relevant for most real systems.  
To show this, we derive explicit closed-form solutions for the full probability distribution of the final cascade size. 
We focus on two topological limit cases, the complete network representing a dense network with a very narrow degree distribution, and the star network representing a sparse network with a inhomogeneous degree distribution. 
Those topologies are of great interest, as they either minimize or maximize the average cascade size and are common motifs in many real world networks.
\end{abstract}
\begin{document}

\flushbottom
\maketitle
\thispagestyle{empty}

\section*{Introduction}
Systemic risk is defined as the risk that a large fraction, $\rho \to 1$, of a system fails. 
This can happen because extreme events shock the system \cite{Tessone2012} or because the failures of a few system elements cause new failures.
This propagation results in failure cascades that can eventually reach the size of the system. 
In the aftermath of the financial crisis in 2008/2009, systemic risks induced by such cascades received attention especially in the context of financial contagion of bank defaults \cite{Gai2010,Haldane2011,delli2012liaisons}. 
Also the increasing complexity of economic value chains \cite{Schweitzer2009c} raised awareness for similar effects \cite{Fagiolo}, in particular concerning critical resources \cite{Klimeke1500522} and food \cite{thesis}.
Many of these works are inspired by stability analyses of ecological food webs \cite{Battiston818,Kondoh1388,CIS-9297}.
But also generic models from physics and related areas have been utilized to study failure cascades.
Examples include Ising models \cite{IsingCascade}, models of fiber bundles that break under stress \cite{RevModPhys.82.499}, models of epidemic spreading \cite{RevModPhys.87.925,Heesterbeekaaa4339}, voter models \cite{PhysRevLett.101.018701}, and models of overload failures in power grids or other infrastructure \cite{PhysRevE.66.065102}.
Many of these cascade models have in common that they can be mapped to a generic threshold model for propagating failures on a graph $G$, or a complex network, where nodes represent the system elements and links between nodes their interactions. 

\paragraph{Network model}
Formally $G = (V, E)$ is a constant undirected network consisting of $N = |V|$ nodes, which are elements of the node set $V$. The link set $E$ contains tuples of the form $(i,j)$, each representing a pairwise connection between nodes $i$ and $j$.
Network topologies vary between Erd\"os-R\'enyi random networks, where the number of links per node can be described by an average degree, and scale-free networks, where the number of links vastly differs between core nodes, or hubs with very many links, and peripheral nodes with very few links. 
In this paper, we discuss limit cases of these two topologies, i.e. the complete network, where each node has the same number of links, $N-1$, and the star network, where the center node has $N-1$ links, but the peripheral nodes have only one link to the center. 
Both are recurring motifs of real world networks and an approximation of prevailing core periphery structures.
In particular, financial and economic networks often have densely connected core nodes with peripheral nodes loosely connected to them \cite{Trivik,ReviewFinanceContagion}. 
Most importantly, for many models, one of the two, the complete network or the star, minimizes while the other maximizes average systemic risk \cite{Acemoglu,NBERw20931}, when systemic risk is measured by the cascade size, i.e. the fraction of failed nodes in the network.

\paragraph{Cascade processes}
A failure of a node $i$ at time $t$ is indicated by its binary state variable $s_{i}(t) = 1$. 
Otherwise, it is healthy or functional, $s_{i}(t)=0$, and can switch to failed in each of the discrete time steps $t = 0, \ldots, T$.
For simplicity, we exclude the possibility to recover, i.e. node states can only change from $0$ to $1$ but not vice versa.
Each node state $s_i(t)$ is characterized by two node variables, a threshold $\theta_i$ and the load $\lambda_i(t)$. 
$\theta_{i}$ summarizes the ``robustness'' of a node, its ability to withstand shocks. 
It is assumed to stay fixed over time.
Individual thresholds are initially drawn independently at random from an arbitrary cumulative distribution function (cdf) $F(\theta)$, often also called the response function.
A node $i$ fails at time $t+1$, if its load exceeds its threshold, i.e. $s_{i}(t+1)=\Theta[\lambda_{i}(t)-\theta_{i}]$, where $\Theta[x]$ is the Heaviside function with $\Theta[x]=1$ if $x\geq 0$ and $\Theta[x]=0$ otherwise.  
The load $\lambda_{i}(t)$ can change because of interaction with neighboring nodes involving a load distribution process. 
This basically describes how a failure can propagate along a link if two nodes interact, because a load increase can result in $\lambda_{i}(t)\geq \theta_{i}$.
Initially, all nodes with  $\theta_{i} \leq \lambda_{0}$ fail, where $\lambda_{i}(0)=\lambda_0$ denotes the initial load of a node.

This allows us to define the size of a cascade at time $t$ as the fraction of failed nodes in the network: 
$\rho(t) = 1/N \sum^{N}_{i=1} s_i(t)$. 
A cascade maximally lasts $T=N-1$ time steps because at least one node needs to fail at each time step to keep the process going.
The final size of the cascade, $\rho=\rho(T)$, i.e. the final fraction of failed nodes, is a common measure for systemic risk. 
While the cascade dynamics itself is deterministic, $\rho$ is a random variable, as it depends on the thresholds that are drawn initially independently at random according to the distribution $F(\theta)$. 
This randomness can model uncertainty (e.g.g incomplete data or information) or changing exposures of nodes, for instance, due to fluctuating markets or aging system components in an engineering setting.
Based on combinatorial arguments, in the materials and methods section and the SI, we derive a closed-form solution for the probability distribution of $\rho$ on complete graphs, $p\left(\rho_{\triangle}\right))$, and stars, $p\left(\rho_{\star}\right))$, of finite arbitrary size $N$.

\paragraph{Three paradigmatic load distribution mechanisms} 
We discuss three exemplary load distribution mechanisms well established in different fields. 
We refer to the first one \cite{Watts2002SimpleModelof} as exposure diversification (ED). 
It means that a node is exposed equally to the risk that any of its neighbors fail. %, which may cause in impact on the node itself. 
The more neighbors, the better for hedging against such exposures (in low systemic risk regimes) \cite{Burkholz2015}. 
A node carries simply the fraction of its failed neighbors as load.
Thus, in complete networks each functional node carries the load $\lambda[k] = k/(N-1)$ when $k$ nodes have failed.

The ED mechanism has not only attracted theoretical interest \cite{Gleeson2007,Payne2009,Hurd2013,BurkholzMultiplex,BurkholzCorrelations}, it is
also used to explain opinion formation \cite{watts2007influentials} and 
financial contagion \cite{Gai2010,Battiston2012a,Amini2010}, where random thresholds correspond to fluctuating exposures between banks and changing capital buffers. 
The randomness can also be interpreted as model uncertainty in such complex systems \cite{BattistonPNAS}.

The second mechanism, denoted as  damage diversification (DD) \cite{Lorenz2009b,Burkholz2015}, assumes that a failing node distributes its total load $1$ equally among its network neighbors. 
Consequently, DD coincides with ED for complete networks. 

The third mechanism defines the prominent fiber bundle model \cite{LRD,RevModPhys.82.499} in material science. 
A force, applied to a bundle of fibers, is modeled by a load $\lambda_0$ carried by each node in a network, while a node's threshold determines the strength of a fiber. 
When a fiber $i$ breaks, it distributes its full load $\lambda_i(t)$ equally to its still functional network neighbors, which become connected afterwards. 
On complete networks, this mechanism is also known as global load sharing rule and translates into $\lambda[k] = \lambda_0 + k\lambda_0/(N-k)$ that each functional node carries when $k$ nodes have failed. 
For the special and simplifying case of uniformly distributed thresholds, a formula for the cascade size distribution is known already \cite{PES:262284}. 
With our new derivations, now arbitrary threshold distributions and thus different fiber materials can be studied.  
For the star network, we focus on a local load sharing variant where no new edges are created \cite{moreno2002instability,BurkholzLRD}

\section*{Results}
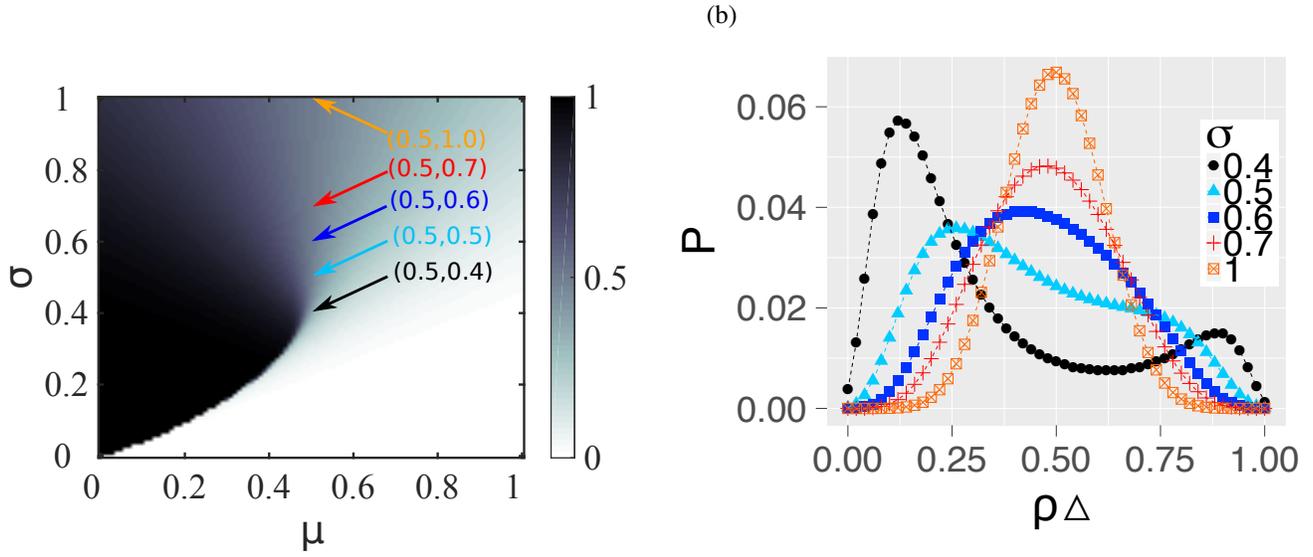
\begin{figure}[t]
 \flushright
 \includegraphics[width=0.49\textwidth]{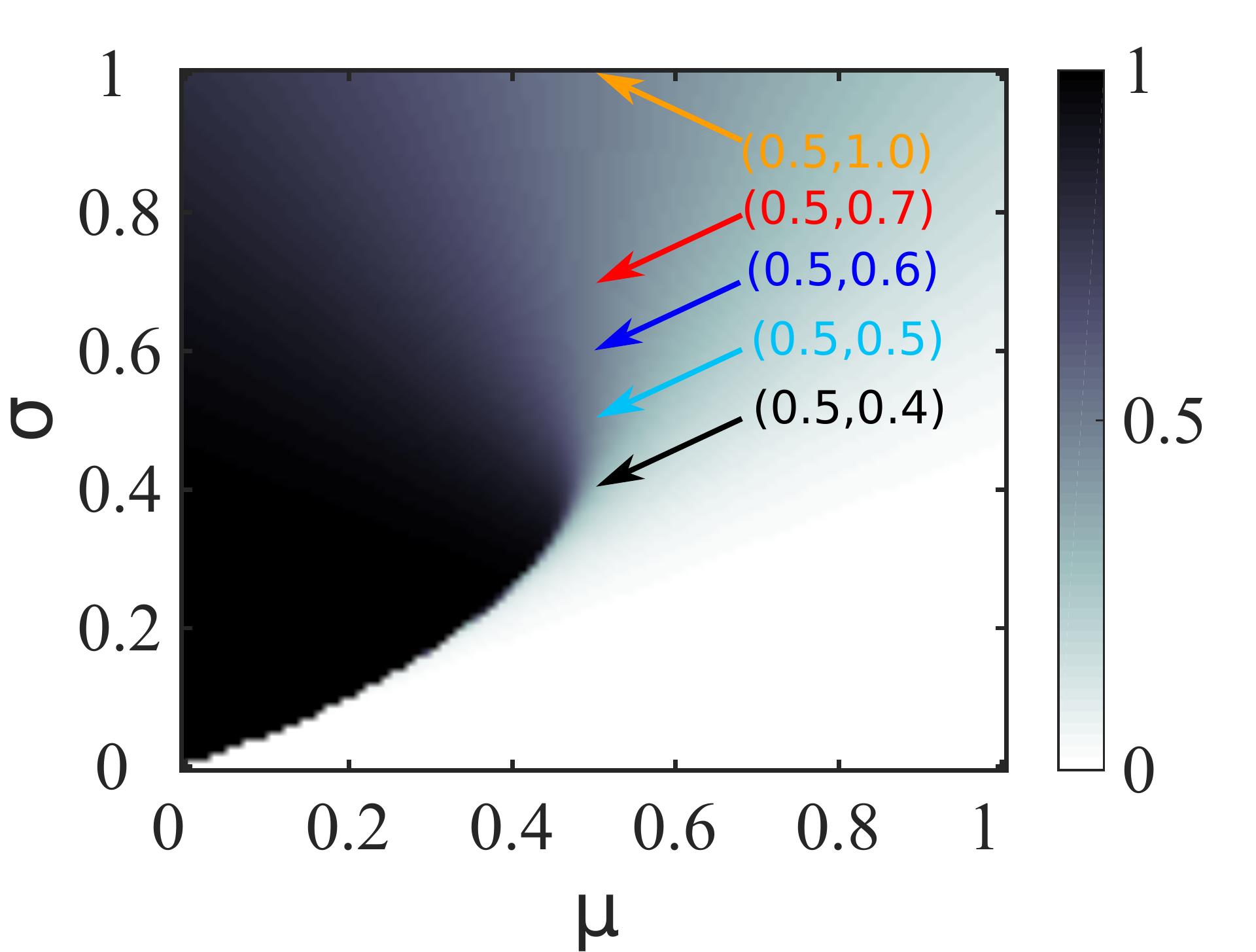}
   \includegraphics[width=0.49\textwidth]{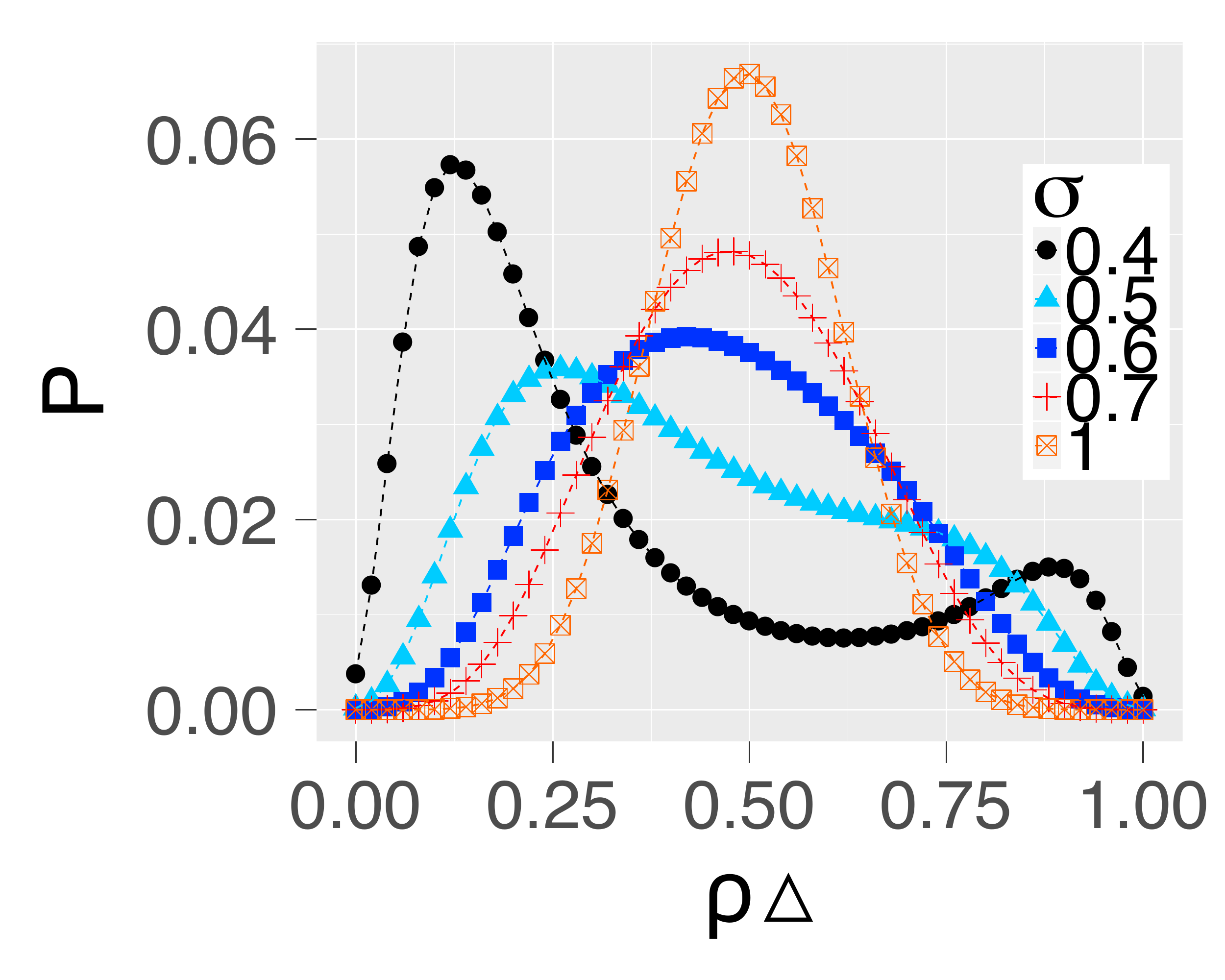}
   \vspace{5mm}
  \begin{picture}(0,0)
    \put(-230,400){(a)}
   \put(-230,200){(b)}
  \end{picture}
\caption{ED/DD mechanism on complete network and normally distributed thresholds, i.e $F \sim \mathcal{N}\left(\mu,\sigma^2\right)$, with mean $\mu$ and standard deviation $\sigma$. (a) (Average) final fraction of failed nodes $\rho_{\triangle}$ (color coded) on infinitely large network. The darker the color, the higher is $\rho_{\triangle}$.  (b) Distribution of $\rho_{\triangle}$ on finite network of size $N=50$ with respect to different $\sigma$. $\mu = 0.5$ is fixed.
}
\label{fig:fc} 
\end{figure}

The only random ingredient that we study is the cumulative threshold distribution $F(\theta)$.
Our derivations apply to an arbitrary choice. 
Yet to be consistent and to allow for comparisons, we discuss only $F\sim \mathcal{N}(\mu + \lambda_0, \sigma^2)$, i.e. a normal distribution with mean $\mu + \lambda_0$ and standard deviation $\sigma$.
For ED and DD, $\lambda_0$ does not influence the results, so we set the initial load to $\lambda_0 = 0$.
For the fiber bundle model, we assume $\lambda_0 = 1$, i.e. a value large enough that we can observe both cascades that span the whole network ($\rho = 1$), and cascades that stop early.  
We are especially interested in parameter regions $(\mu,\sigma)$ that lead to medium average cascade sizes.
This implies that such cascades would not encompass the whole system in case of an \emph{infinitely} large network. 
We then investigate, for such ``safe'' parameter regions, the size of cascades for \emph{finite} networks.

As a reference case, Fig.~\ref{fig:fc}(a) shows the phase diagram for $\rho_{\triangle}$ for infinitely large systems, assuming a complete network and the ED/DD mechanism \cite{Lorenz2009b,Burkholz2015}.
$\rho_{\triangle}$ is taken as a measure of systemic risk.
If the mean value of the threshold distribution, $\mu$, is large, the system is less prone to large failure cascades. 
Taking a small, but fixed value of $\mu$, we recall two important observations, (i) the sharp phase transition between completely safe systems (white, average cascade size zero) and completely failed systems at a critical value of $(\mu,\sigma)$, (ii) the gradual decrease of systemic risk in areas with high $\sigma$. 
This means, the heterogeneity in thresholds plays a crucial role in amplifying, or preventing, systemic risks. 
Therefore, we first study the impact of increasing heterogeneity $\sigma$ on failure cascades in \emph{finite} networks.
Fig.~\ref{fig:fc}(b) shows the broad distribution of the final cascade size, spanning values between 0 and 1. 
Only for large values of $\sigma$ we find distributions that match our expectations derived from the behavior of infinite systems, i.e. they are unimodal and symmetric.
As $\sigma$ decreases, however, the distributions become asymmetric and broader and even bi-modal. 
Only very close to a phase transition, broader distributions in finite systems are expected by theoretical physicists and, usually, explicit formulas for such distributions are unknown.  
But even for parameters further away from the phase transition in Fig.~\ref{fig:fc}(a), the average cascade size is meaningless to characterize the system. 
Instead, we can expect either small cascades or, as a new observation, very big ones that are likely to destroy the whole system and are relevant for systemic risk estimations. 
Counter-intuitively, the smaller the threshold heterogeneity, the the less predictive the system behavior becomes.

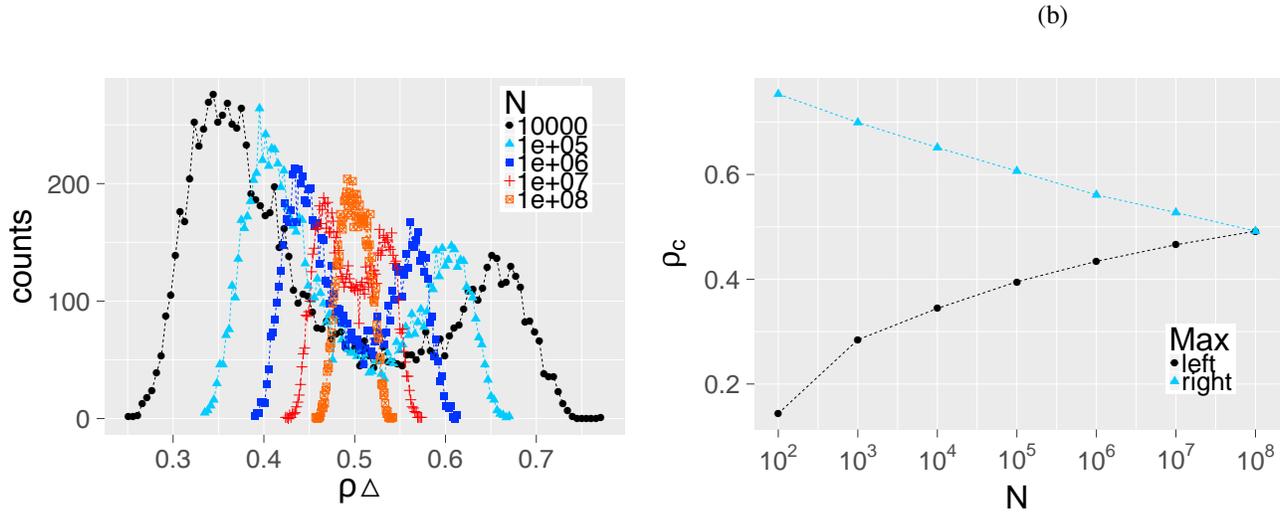
\begin{figure}[t]
 \centering
\includegraphics[width=0.49\textwidth]{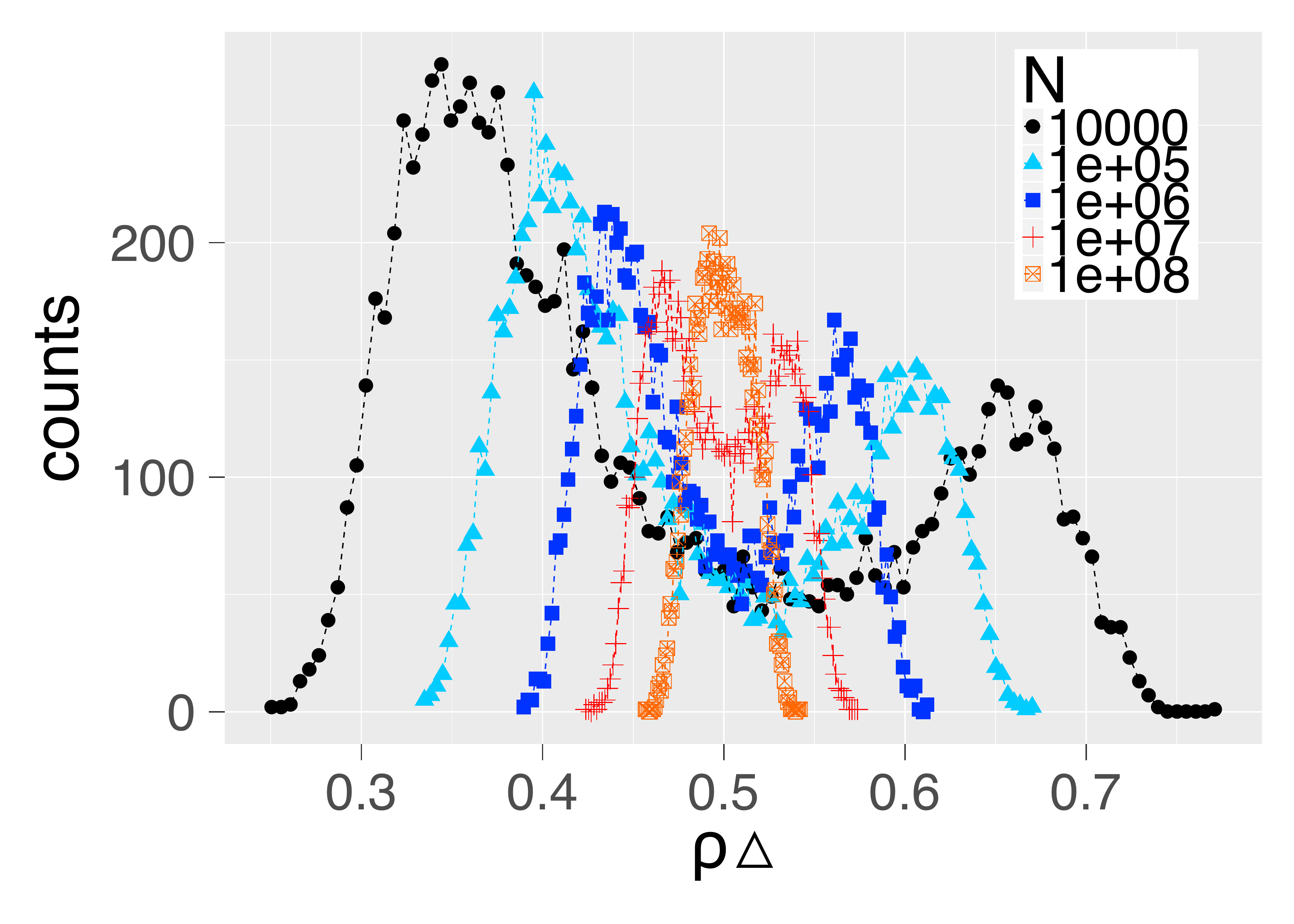}
\includegraphics[width=0.49\textwidth]{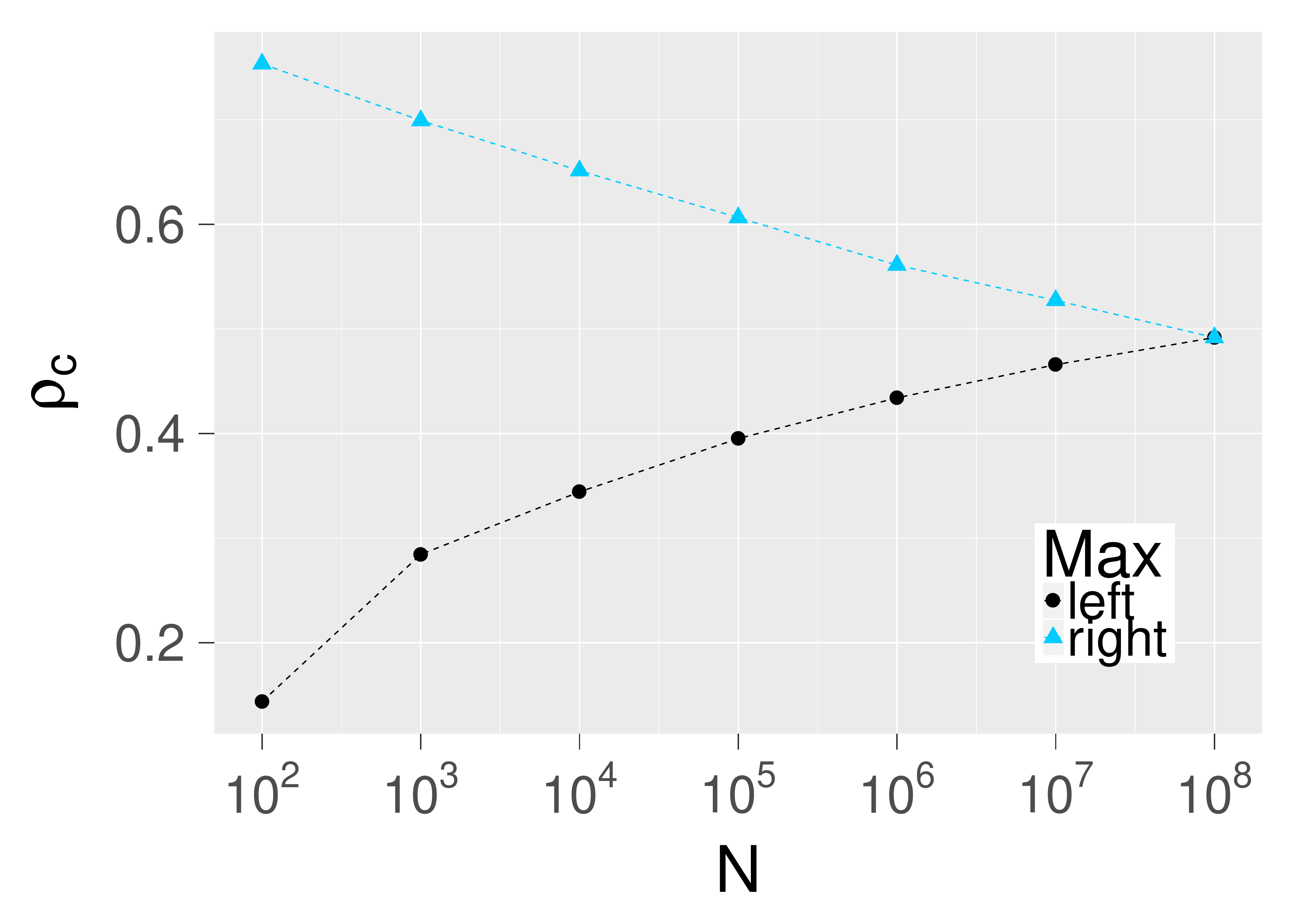}
\begin{picture}(0,0)
    \put(-100,370){(a)}
   \put(-100,190){(b)}
   \end{picture}
 \caption{(a) Final cascade size distribution for complete networks of several sizes $N$ obtained via $10^4$ independent simulations of the  ED mechanism, i.e. draws from $P\left(\rho_{\triangle}\right)$. The thresholds are independently normally distributed with mean $\mu = 0.5$ and standard deviation $\sigma = 0.4$ ($F \sim \mathcal{N}\left(0.5,0.4^2\right)$). (b) Position of the two local maxima of the distributions in (a) with respect to the network size $N$.}
 \label{fig:convergenceFinite}
\end{figure}

This strong bi-modality is a new observation that only applies to \emph{finite} systems and is not captured by risk approximations for infinite systems. 
To further investigate how this behavior depends on the system size $N$, we simulate, for parameters $(\mu,\sigma) = (0.5,0.4)$, the distribution of final cascade sizes. 
Monte Carlo simulations are necessary because, for systems up to $N=10^{8}$, the products of large binomial terms and small probabilities that appear in our exact derivations become unfeasible to solve numerically. 
Fig.~\ref{fig:convergenceFinite} shows the results. 
We see that the bi-modality persists even for networks of size $N=10^7$. 
So, the question whether networks of that size can be reasonably described by approximations for infinite networks, is clearly answered by ``no''.
On the other hand, real economic networks which are prone to failure cascades are often of the order 10$^{4}$-10$^{6}$, for example ownership networks or collaboration networks of firms \cite{Schweitzer2009c}.
Hence, it is of importance to correctly reflect finite-size effects in risk estimations of such systems. 

\begin{figure}[t]%[h]
 \centering
  \includegraphics[width=0.49\textwidth]{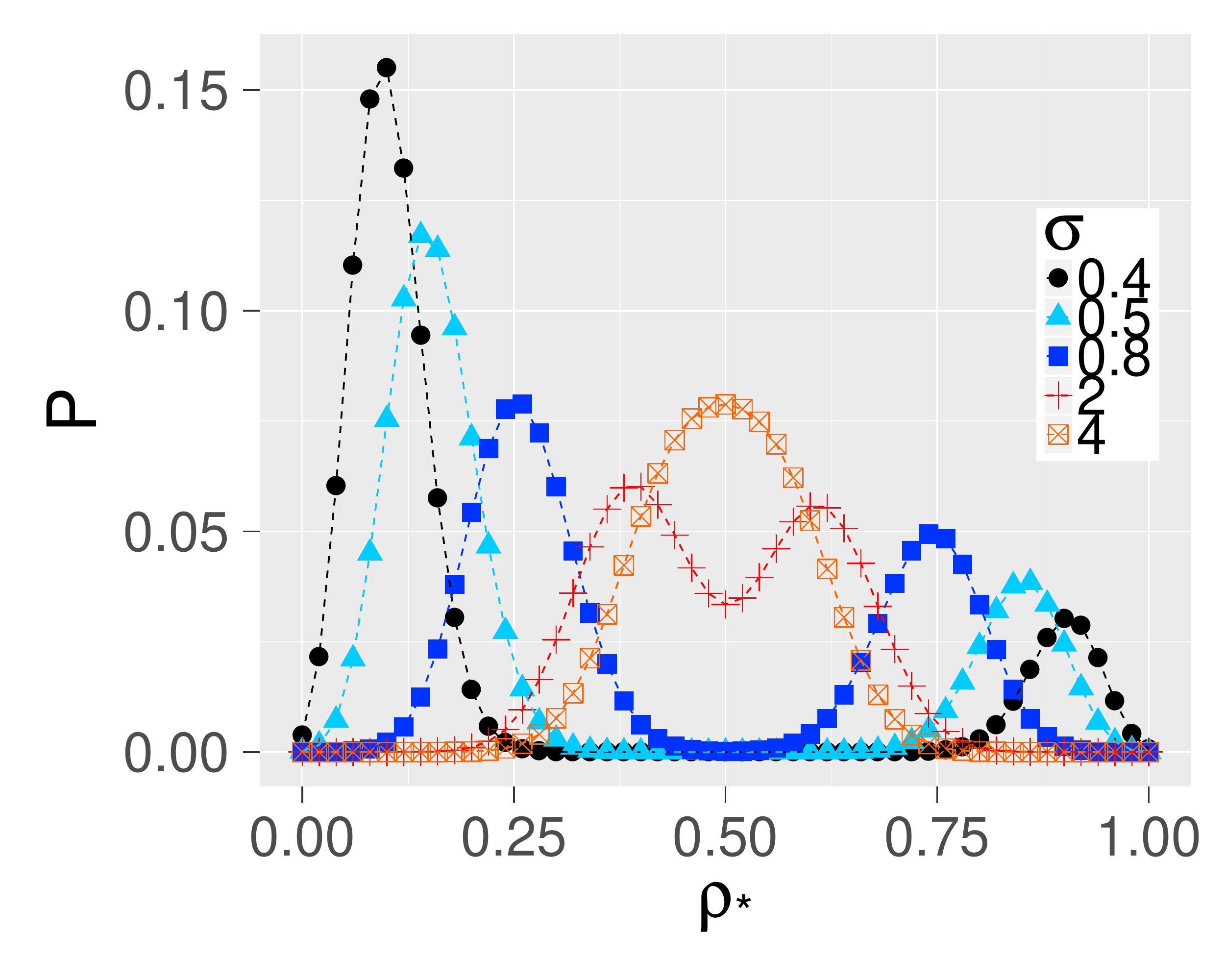}
 \includegraphics[width=0.49\textwidth]{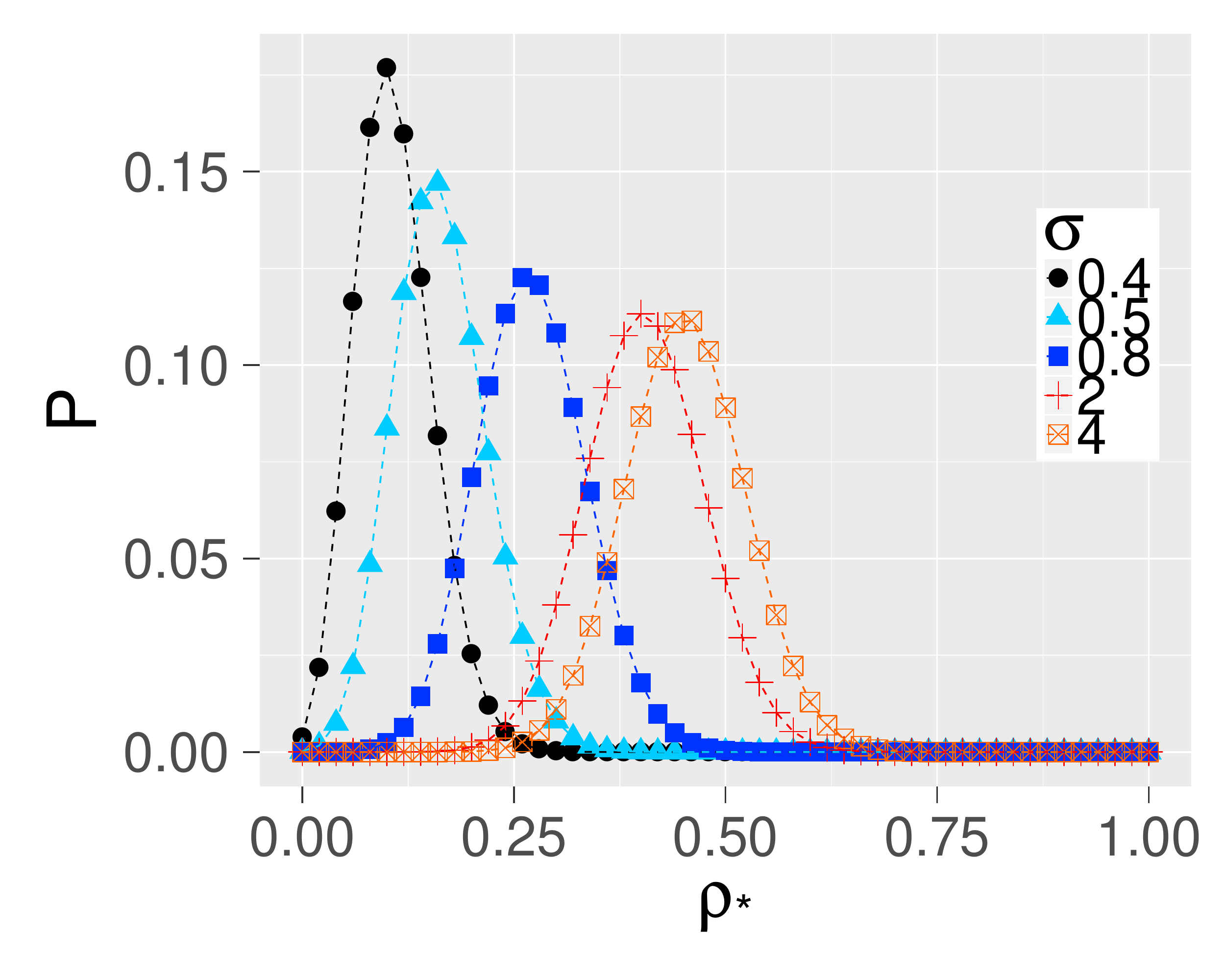}
\begin{picture}(0,0)
     \put(-100,410){(a)}
   \put(-100,210){(b)}
   \end{picture}
\caption{Probability distribution of the final fraction of failed nodes $\rho_{\star}$ on a star network with $N=50$ nodes for the (a) ED mechanism and (b) DD mechanism. The thresholds are normally distributed with mean $\mu=0.5$ and standard deviation $\sigma$ (i.e. $F \sim \mathcal{N}\left(0.5,\sigma^2\right)$.}
\label{fig:stars}
\end{figure}

Fig.~\ref{fig:stars} shows the corresponding cascade size distributions for the star-shaped networks. 
Here, we have to distinguish the ED and the DD mechanisms of load distribution, which coincide only for complete networks. 
Fig.~\ref{fig:stars}(a) shows the results for the ED mechanism. Again, we observe bi-modality in the distribution of cascade sizes that vanishes only for large threshold heterogeneity $\sigma$.  
%Again, for increasing  $\sigma$ the two peaks of the distribution merge to a single one.
The bi-modality has a clear interpretation with respect to the central node.
In smaller cascades, the center does not fail, which explains the peak at low values of $\rho_{\star}$.
But if the center fails, it triggers a significant number of further failures of leaf nodes that explains the right peak in the distribution.
This bi-modality is so fundamental that it does not even vanish in the limit $N \rightarrow \infty$, as we show in the materials and methods section.

The results for the DD mechanism presented in Fig.~\ref{fig:stars}(b) show that, in contrast to the ED mechanism, the distributions of the final cascade size are unimodal, even for smaller $\sigma$.
Again, this can be explained with the role of the central node. 
Although the center fails with high probability, with the DD mechanism, its failure distributes only a small amount of load equally to the (large number) of leaf nodes. 
Hence, this does not cause substantial additional failures.
As in infinitely large systems \cite{Burkholz2015}, in the presence of hubs, failure cascades involving the DD mechanism expose the system to a smaller failure risk than cascades involving the ED mechanism - at the expense that hubs usually fail.

\begin{figure}[t]
 \centering
\includegraphics[width=0.49\textwidth]{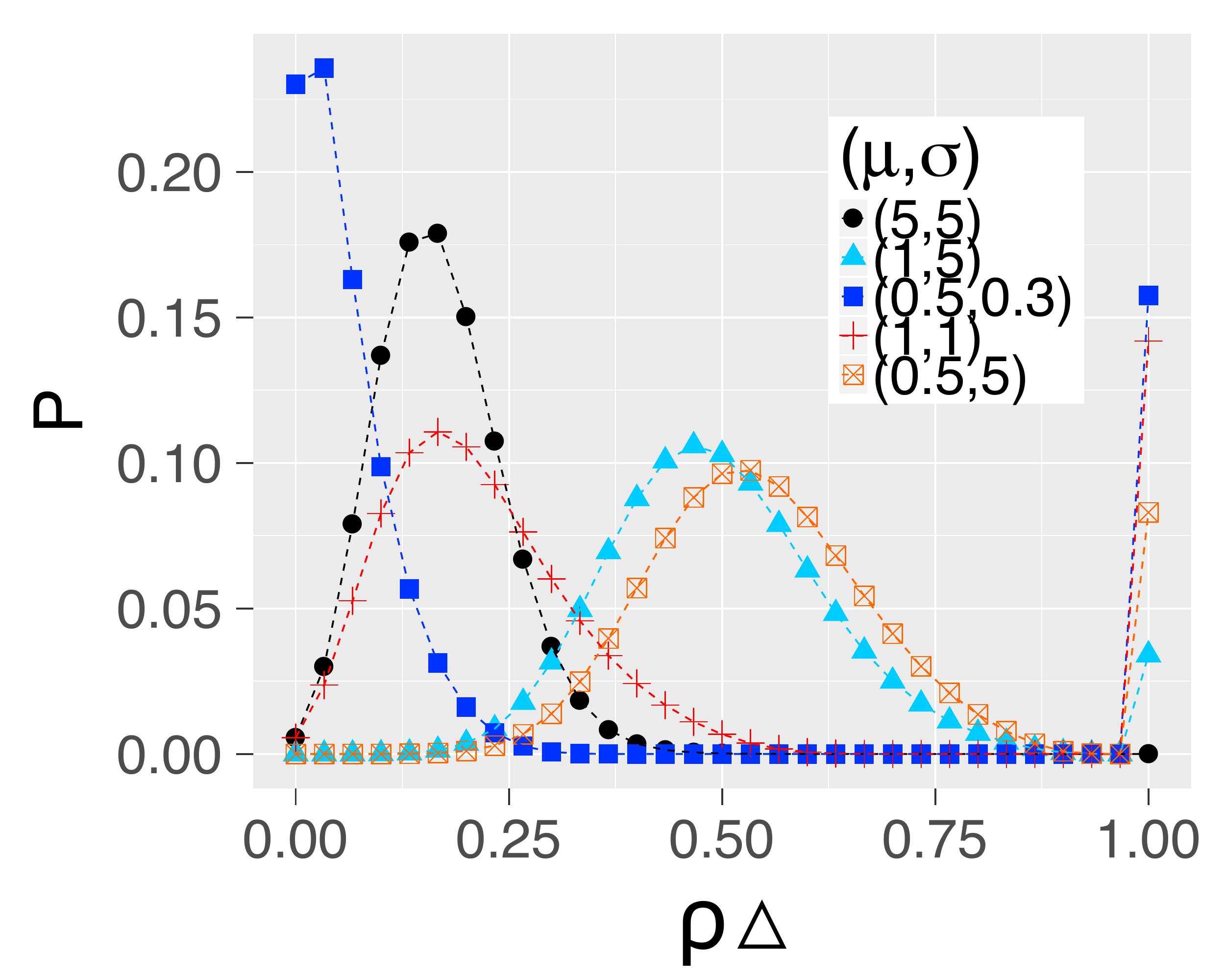}
\includegraphics[width=0.49\textwidth]{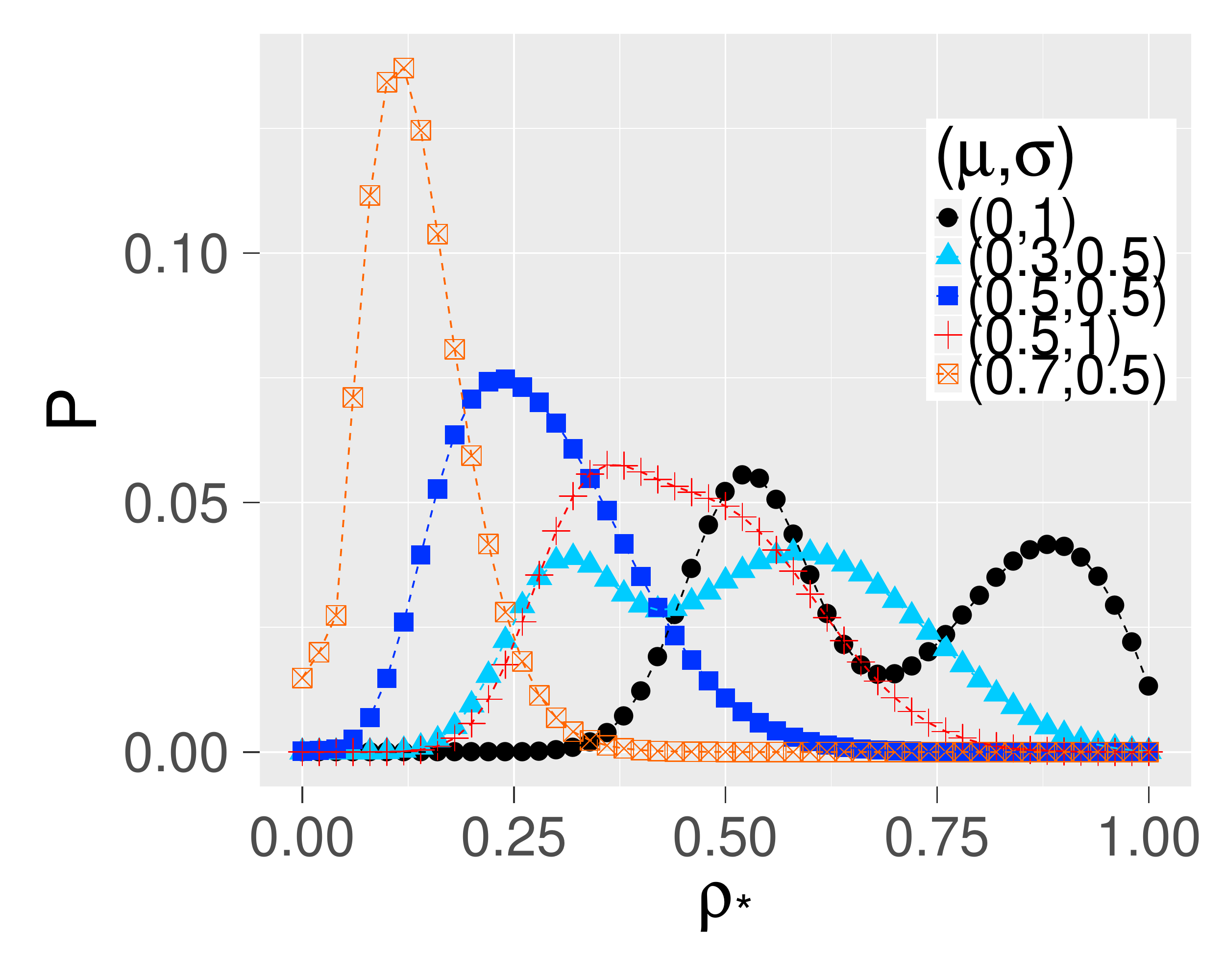}
\begin{picture}(0,0)
    \put(-100,410){(a)}
   \put(-100,210){(b)}
   \end{picture}
 \caption{Fiber bundle model with initial load $\lambda_0 = 1$. Probability distribution of the final fraction of failed nodes with respect to normally distributed thresholds, i.e. $F \sim \mathcal{N}\left(\mu + \lambda_0,\sigma^2\right)$, on (a) complete network of size $N=30$  and (b) star of size $N=50$.}
 \label{fig:LRD}
\end{figure}

Eventually, we also discuss the third load distribution mechanism used in the fiber bundle model. 
Because of the additional load $\lambda_0$, more load is distributed after a cascade started.
As a result, in complete networks large cascades amplify to a full system break-down  shown in the very right peak at $\rho_{\triangle} = 1$ in  Fig.~\ref{fig:LRD}(a). 
This is an extreme case of the bi-modal distribution, where also small cascades can occur with high probability.
As Fig.~\ref{fig:LRD}(b) shows, in star-like networks the range of probable cascade sizes is usually bigger.
Dependent on the $(\mu,\sigma)$ values of the threshold distribution, bi-modal cascade-size distributions can still occur. 
The peaks of the cascade-size distribution can be explained again by the survival or the failure of the central node.  
In comparison with complete networks, these peaks are often close together. 
Further, the lower connectivity of the star-like networks reduces the risk of large cascades.

\section*{Discussion}

Our driving question was whether systemic risk in finite systems can be reasonably estimated from approximations valid for infinite systems. 
We have to negate this question for two reasons. First, taking the final size of failure cascades as a risk measure, we could demonstrate, for the first time, that the distribution of this quantity changes fundamentally in finite systems. 
While a unimodal and narrow distribution, typical for infinite systems, would allow to use the average cascade size as a reasonable measure, we could show that for finite systems we often have a very broad and even bi-modal distribution. 
Hence, an estimated average does not quantify the real risk. 
Instead, small cascades but also extreme larger failure cascades become more likely.
Even though a complete system break-down is unlikely as we show (which may be expected for infinite, but not for finite systems), large cascades in finite systems are much more frequent than expected so far. 
Secondly, we could also estimate the range of sizes for which such finite size effects play a role.
We found that they are important for system sizes up to $10^{7}$, i.e. these systems are quite large, even though finite. 
This demonstrates that corrections are needed not just for small systems of a few to a few hundred nodes.
Instead, they are already necessary for all real-world systems, up to a hundred million nodes. 

In addition to these insights, we were able to derive novel explicit closed-form solutions to calculate the full distribution of the final cascade size, for two exemplary topologies. 
Our results are unique, as we derive exact solutions that apply to networks of arbitrary finite size $N$, a wide range of load distribution mechanisms, and arbitrary threshold distribution to characterize the robustness of the nodes. 

In light of the combinatorial complexity of the problem, such results are rare and crucially depend on the high symmetry of the considered networks. 
Complete networks and stars are of particular interest, as they appear to be systemic risk minimizing or maximizing when only the average cascade size is considered \cite{Acemoglu,NBERw20931}.
Intuitively, this is reasonable, as the complete graph represents the limit case of perfect risk diversification, where the risk is shared between all system components. 
The star network can be less prone to the amplification of failures in the course of a cascade, in particular when its central node is very robust. {For that reason, central counter parties are often favored to reduce systemic risk.}   
Yet, we have to reconsider our assessment of robust finite network topologies and take the probability of large system break-downs into account, as these are relevant for risk averse decision makers. 

Many further real world networks, especially economic networks, show a core-periphery structure \cite{Trivik,ReviewFinanceContagion} with densely connected core nodes and  peripheral nodes loosely connected to the core. 
Thus, their topology can be understood as a superposition of the sample topologies discussed here, the complete network and the star network. 
Sparse topologies, with the star network as a paragon, occur frequently in real world networks \cite{Barabasi} and play an important role in the spreading of failures \cite{Dorogovtsev}. 
These systems usually have sizes that require to consider the full cascade size distribution for the estimation of systemic risk, as we propose with our solutions. 

Beyond their elegance, such closed-form formulas have the advantage that they enable explicit optimization strategies for control of the involved parameters. For instance, we can derive derivatives with respect to those parameters that inform gradient based optimization methods. 
This allows to design robust systems, by controlling e.g. the broadness  or the convergence of the cascade size distribution. 
It can be also applied in restoration strategies for failed systems \cite{restoration}.  

Yet, we cannot expect to derive such analytic results for arbitrary network topologies. In this case, we have to rely on Monte Carlo simulations to estimate the broadness and shape of the cascade size distribution. 
We hope to inspire systemic risk analysts to base their insights and decisions on risk measures that can capture such broadness and multi-modality.

\section*{Methods}

\subsection*{Derivation of cascade size distributions} 
In finite networks, the final cascade size $\rho$ takes discrete values in the set $\left\{0, 1/N, 2/N, \ldots, 1\right\}$. 
To determine the distributions $P(\rho_{ \triangle})$ and $P(\rho_{\star})$, we can calculate the probabilities for each event $\mathbb{P}\left( \rho_{\triangle} = k/N \right)$ and 
 $\mathbb{P}\left( \rho_{\star} = k/N \right)$, where $k$ denotes the number of failed nodes at the end of the cascading process.

\subsubsection*{Complete networks}

For complete networks, we can decompose the probability of $k$ final failures into the product of the probability that $N-k$ out of $N$ nodes do not fail and the probability $p_k$ that exactly $k$ nodes fail
\begin{align}
    P\left(\rho_{\triangle} = \frac{k}{N} \right) = \binom{N}{k} \left(1- F \left(\lambda[k] \right)\right)^{N-k} p_k.
\label{eq:1} 
\end{align}

Recall that $\lambda[k]$ denotes the load that a node with $k$ failed neighbors carries so that $F \left(\lambda[k] \right)$ is the probability that this load exceeds the node's threshold. 
With the help of the inclusion-exclusion principle \cite{inclexcl}, $p_k$ can be expressed as
 \begin{align}\label{eq:pkfc}
    p_k = \sum^{k-1}_{l=0} (-1)^{k+l+1} \binom{k}{l} F\left(\lambda[l] \right)^{k-l} p_l.
 \end{align}
The details of this derivation and proofs are provided in the supplementary information.
Apparently, our derivations apply to an even larger class of models than threshold models. 
The failure probability of a node, here $F \left(\lambda[k] \right)$, can be any function of the number of its failed neighbors.

\subsubsection*{Star networks}
The central node with degree $d = N-1$ has a prominent role, as it is the only node that can distribute accumulated load received by other nodes.
We allow it to have a different threshold cdf $F_c$ and load function $\lambda_c$ than the remaining $N-1$ leaf nodes, which have degree $1$, threshold cdf $F$ and carry loads of the form $\lambda_{r}$.  
A leaf node can fail initially with probability $F\left(\lambda_{r}[0]\right)$ (with $0$ neighboring failures) or, if the center has failed before, it can carry the load $\lambda_{r}[j,l]$, which depends on two additional variables that are determined by the history of the center: $j$ and $l$.  
$j$ denotes the number of failed leaf nodes that have distributed load to the center, before the center has failed. 
$l$ indicates the number of nodes among which the accumulated load of the center is shared (when the center fails). 
We simply add the probabilities for all possible cases and obtain: 
%\begin{widetext}
\begin{align}\label{eq:star}
  \begin{split}
 &  P\left(\rho_{\star} = \frac{k}{N} \right)    =  \left(1- F_c\left(\lambda_c[k]\right)\right)  \binom{N-1}{k}  F\left(\lambda_{r}[0]\right)^{k}  \left(1- F\left(\lambda_{r}[0]\right)\right)^{N-1-k}\\ 
 & +  F_c(\lambda_c[0])  \binom{N-1}{k-1}  \sum^{k-1}_{j=0} \binom{k-1}{j} F\left(\lambda_{r}[0]\right)^{j} \left(F\left(\lambda_{r}[j,N-1-j]\right)- F\left(\lambda_{r}[0]\right)\right)^{k-1-j}\\
 & \times \left(1- F\left(\lambda_{r}[j,N-1-j]\right)\right)^{N-k} +  \binom{N-1}{k-1}  \sum^{k-1}_{j = 1}  \binom{k-1}{j}  F \left(\lambda_{r}[0]\right)^{j} \left(F_c\left(\lambda_c[j]\right) - F_c \left(\lambda[0]\right)\right) \\
& \times  \left(F\left(\lambda_{r}[j,N-1-j]\right) - F \left(\lambda_{r}[0]\right)\right)^{k-1-j} \left(1- F\left(\lambda_{r}[j,N-1-j]\right)\right)^{N-k} .
\end{split}
\end{align}
%\end{widetext}
The first summand considers the case when the center does not fail, while the second term adds the probability for the case when the center fails initially. 
Then, each of the other $k-1$ failures of leaves can either occur initially or because of a load distribution by the center. 
The size of this load might depend on the number of nodes $l$ that fail initially together with the center, since these nodes cannot receive load after the failure of the center. 
The index $j$ in the third term finally takes the events into account when $j$ leaves have failed before the center.

Remarkably, this distribution converges for $N\rightarrow \infty$ to a discrete distribution with non-zero probability at three events, if $\lambda_c[k]$ and $\lambda_{r}[j,l]$ only depend on fractions of nodes ($\lambda_c[k]=\lambda_c[k/N], \lambda_{r}[j,l] = \lambda_{r}[j/N,l/N]$).
In an infinitely large star, exactly $F(\lambda_0)$ of the leaves fail initially.
With probability $F_c(\lambda_0)$, the center node fails as well initially so that $\rho_\star = F(\lambda_{r}[0,1-F(\lambda_0)])$ of the nodes fail in total.  
With probability $F_c(\lambda_c[F(\lambda_0)]) - F_c(\lambda_0)$, the center fails after the initial failures, which leads to $\rho_\star = F(\lambda_{r}[F(\lambda_0),1-F(\lambda_0)])$.
Or with probability $1-F_c(\lambda_c[F(\lambda_0)])$, the center does not fail and we have $\rho_\star = F(\lambda_0)$.
In case of the ED model, two cases collapse to one because $\lambda_{r}[0,1-F(\lambda_0)] = \lambda_{r}[F(\lambda_0),1-F(\lambda_0)] = 1$. 
Thus, the observed bi-modality of the distributions in Fig.~\ref{fig:stars}~(a) persists for increasing network size $N$.

\subsection*{Exemplary load distribution mechanisms}
We have discussed three basic load distribution mechanisms corresponding to the exposure diversification (ED) model, the damage diversification (DD) model and a fiber bundle model.
In the following, we specify how these models define the loads used in the derivations above.

\subsubsection*{ED load redistribution}
In the complete graph, each load carries the load $\lambda[k] = k/(N-1)$.
For the star network, $\lambda_c[k] = k/(N-1)$ holds for the center node and $\lambda_r[0]=0$, $\lambda_r[j,l] = 1$ for the leaf nodes. 

\subsubsection*{DD load redistribution}
In the complete graph, each load carries the load $\lambda[k] = k/(N-1)$. 
On the star network, DD corresponds to the choice $\lambda_c[k] = k$, $\lambda_r[0]=0$, $\lambda_r[j,l] = 1/(N-1)$.

\subsubsection*{Fiber bundle model load redistribution}
On complete networks, the total load $\lambda_0 k$ has been redistributed so that each functional node carries $\lambda[k] = \lambda_0 + k\lambda_0/(N-k)$.
For stars, we have $\lambda_c[k] = (k+1)\lambda_0$, $\lambda_r[0]=\lambda_0$, $\lambda_r[j,l] = (j+1)\lambda_0/l$. 

For finite networks of arbitrary size $N$, we have obtained explicit closed forms of the full probability distribution of final cascade sizes for two prominent topologies. 
This is a major achievement that allows us to calculate the systemic risk in a highly efficient manner. More detailed information can be found in the supplementary information.

\bibliography{BGS_SystRisk_CLU}

\section*{Acknowledgements}

RB acknowledges support by the ETH48 project. FS acknowledges support by the EU-FET project MULTIPLEX 317532.

\section*{Author contributions statement}

R.B., F.S., H.J.H. designed the research. R.B. derived the formulas for the cascade size distribution, implemented them, and performed corresponding simulations. All authors discussed the results and wrote the paper. 

\section*{Additional information}

The authors declare no competing financial interests.

\end{document}